\documentclass[12pt]{iopart}
\usepackage{graphicx}
\usepackage{epstopdf}
\begin{document}
\title[Nonlinear reflection by an array of double silicon elements]{Analysis of terahertz wave nonlinear reflection by an array of double silicon elements placed on a metal substrate}
\author{N V Sydorchuk$^1$, S L Prosvirnin$^{1,2}$, Y Fan$^3$, F Zhang$^3$}
\address{$^1$ Institute of Radio Astronomy, National Academy of Sciences of Ukraine, Kharkiv 61002, Ukraine}
\address{$^2$ Karazin Kharkiv National University, Kharkiv 61077, Ukraine}
\address{$^3$ School of Science, Northwestern Polytechnical University,  Xi'an 710129, China}
\eads{ryazan@rian.kharkov.ua, prosvirn@rian.kharkov.ua, phyfan@nwpu.edu.cn, fuli.zhang@nwpu.edu.cn}

\begin{abstract}
We demonstrate results of a simulation of nonlinear reflection of intensive terahertz radiation from a silicon-on-metal metasurface involving an excitation of a high-Q trapped mode resonance.  Conditions are presented to observe effects of bistability and hysteresis of nonlinear reflection.
\end{abstract}

\noindent{\it Periodic structure, reflective metasurface, localized field, resonance\/}

\submitto{\JPD}
\maketitle

\section{Introduction}

Modern optics technologies open an opportunity to produce resonant selectively reflecting metasurfaces. A metasurface is a planar patterned metal-dielectric or all-dielectric layer placed on a metal substrate.  Thickness of a metasurface is usually very small in comparison with a electromagnetic field wavelength in a free space. However, the periodic patterning of a structure provides conditions for excitation of different kinds of resonances. A metasurface response manifests in reflectivity, absorption, radiation when using laser media, in electromagnetic field boundary values, which may be the same as on a surface of an artificial magnetic wall, and in confining of the intensive electromagnetic field inside the structure. Wavelength resonant selectivity of metasurface reflection is important in applications related to highly  sensitive detection \cite{absorber_for_sensor_2015, multispectral_sensing_2015, sydorchuk-2017}, in particular, of biological materials.

The feature of metasurfaces to confine the intensive electromagnetic field is important to design light controllable devices  \cite{boyd-2008}  using nonlinear properties of materials. Nonlinear media may be include in the device as some elements of the patterned structure  which constitutive parameters have dependance on the field intensity. A bistability and hysteresis are the feature effects of the reflection by a nonlinear metasurface. Due to these effects nonlinear resonant reflecting metasurfaces are quite promised for application as modulators, limiters, switches and other devices of the THz frequency range.

Thus, in this paper we for the first time demonstrate a numerical simulation of nonlinear reflection of intensive terahertz radiation from the silicon-on-metal metasurface involving an excitation of a high-Q trapped mode resonance.

\section{Reflective array design and theoretical methods}

\begin{figure}[]
	\centering
	\includegraphics[width=12.0cm]{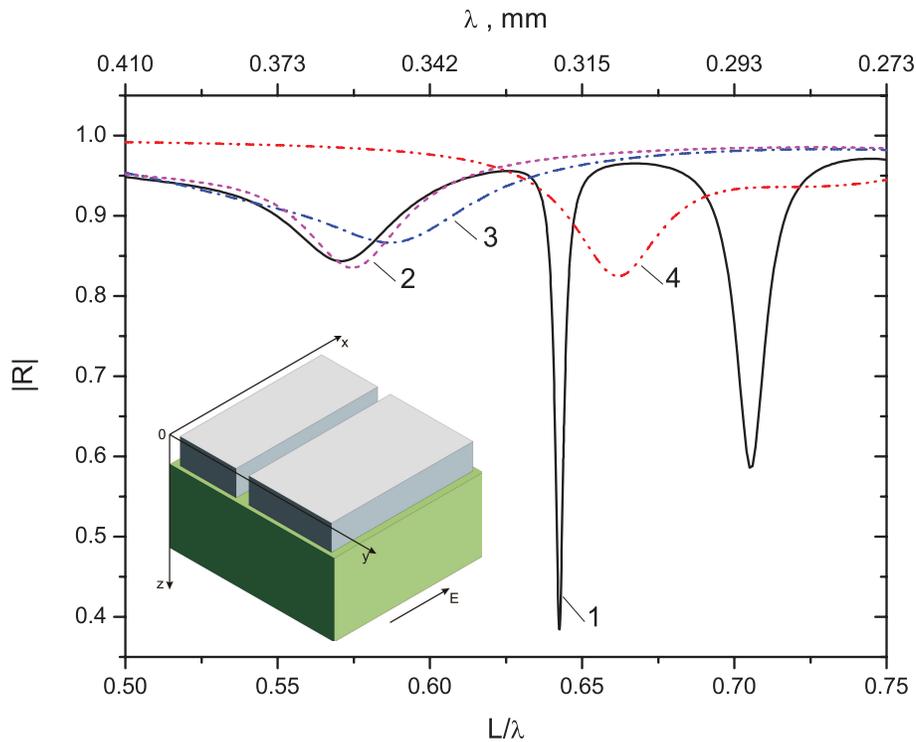}
	\caption{Frequency dependences of normally incident plane wave reflection by metasurfaces with unit cell consisted of two bars of different widths (solid line, 1), two identical bars (dashed line, 2), alone bar the same as wider one of double element structure (point-dashed line, 3), and alone bar the same as narrower one of double element structure (two-points-dashed line, 4). In insert a unit cell of the periodic reflective metasurface consisting of two silicon bars of different widths placed on a copper foil has been shown.}
	\label{RonFrequency}
\end{figure}

A square unit cell of the proposed reflective periodic metasurface is shown in the insert of Fig.~\ref{RonFrequency}. The structure is an array of silicon bars, which are placed on a metallic substrate (a copper foil). The array is periodic in two directions $0x$ and $0y$. Each unit cell contains two silicon bars of the same thickness and length, but of different widths. We have chosen such asymmetric unit cell of the metasurface to provide an excitation of high-Q trapped mode resonance \cite{khardikov-2012-jop}. The pitch of the array along $0x$ and $0y$ axes is $L=205$~mkm.  Length of the bars is 195~mkm. The bars with widths of 50~mkm and 123~mkm were located at a distance of 16~mkm from each other. Thickness of the bars and copper foil is 35~mkm and 100~mkm, respectively.In potential applications, a metal substrate is envisaged as a good heat removed constructive element reducing the time of response on a control variation of incident wave intensity.

For the simulation of the array spectral reflectivity  in a linear regime the rigorous numerical-analytical method and the developed computer code were used. They are based on volumetric integral equations of macroscopic electrodynamics for solving problems of electromagnetic wave scattering by planar multilayer magnetodielectric structures which are periodic in two directions along their plane \cite{sydorchuk2012resonant}. In this approach the electromagnetic field at some point is the sum of the incident wave field and the field scattered by the structure. These representations for the field are converted into integral equations for the equivalent electric and magnetic polarization currents if the observation point is inside the scattering structure. The equations are solved using the  double Floquet–Fourier series expansion, and then the scattered field outside the structure can be calculated. A form of the periodic cell and the values of dielectric permittivity of the structure materials are not limited in the framework of the method. Dielectric permittivity can be complex; its imaginary parts can be positive or negative. The developed numerical code has been tested in a vast number of scattering problems and a comparison with results of different commercial codes and measurements has been made. The tests manifested an excellent accuracy of the developed code and convenience in applied problems of electromagnetics.

In the nonlinear regime, for relative dielectric permittivity  we assume  the Kerr's dependence on the electric field intensity. We use the approximation of monochromatic waves and study the spectral properties of this reflecting metasurface at normal incidence of a plane $x$-polarized electromagnetic wave impinging from the $z<0$ semi-space.  The base of numerical code is consisted of the above mentioned method of volumetric integral equations of macroscopic electrodynamics which was combined with a set of two transcendent equations on the field intensity in the middle points of each bar of the unit cell. The solution of this set of equations was used to determine an actual value of the relative permittivity of bars. The numerical solution of the set was found by the method of simple iterations with using a relaxation.
                                                                            
In the simulations, that results are presented below, we assume the relative dielectric permittivity of $n$-type silicon of bars $\epsilon=\epsilon_l+\chi_3 |E|^2$ where $\epsilon_l=11.70-i0.01$ that is, the electromagnetic material losses were taken into account, and the third-order nonlinear susceptibility is $\chi_{3} = 3.5 \times 10^{-8}$~esu
\cite{narkowicz_2011_Third}. For the dielectric permittivity of the substrate the reference book data \cite{hagemann1975optical} for copper in the terahertz frequency range was used $\epsilon_s=-13700-i159000$.

\section{Linear reflection regime of the metasurface}

The proposed controllable metasurface has a set of sharp resonances clearly manifested as maxims of absorption in the terahertz range of frequencies.  Results of the simulations of electromagnetic wave reflection by the investigated structure are shown in  Fig.~\ref{RonFrequency}. On the upper abscissa axis the corresponding wavelength is presented. It is noticeable that in the case of the bars of different widths there are additional resonances on the reflectance characteristic that are not present in the spectrum of the symmetric structure and the structure with one bar per unit cell.

In order to study the physical origin of the resonances within the metasurface, numerical simulations were made and maps of the internal fields in the studied structure were depictured. In this way we reveal the patterns of distribution of the fields and the places of greatest confinement of electromagnetic power at the resonances.

The first low-frequency resonance (Fig. \ref{RonFrequency}, lines 1-4)  is due to the formation of the electromagnetic field typical of a magnetic dipole parallel to the $0y$ axis with a maximum field in the wide bar inside each unit cell. The electric and magnetic fields reach 23 and 175 conventional units, respectively.

In conditions of the second high-Q resonance next in direction of increasing frequency (Fig. \ref{RonFrequency}, line 1) we have observed opposite directed electric dipoles along each bar, parallel to the $0x$ axis. The maximum values of electric and magnetic fields in the narrow bar are respectively 63 and 830 conventional units. This resonance is a most promised one for investigation of nonlinear response and applications due to its high quality factor and such field distribution along unit cell that a maximum of the electric field intensity is inside of the bars. These two conditions provide advantageous opportunities for electromagnetic field coupling with the nonlinear silicon of bars.  The peculiarity of the resonance is provided by excitation of a trapped mode field distribution inside an unit cell. Its relative frequency is 0.6426. 

Next third resonance (Fig. \ref{RonFrequency}, line 1) is accompanied by the appearance of two magnetic dipoles in one direction, parallel to the $0y$ axis, in each bar and oppositely directed in the wide and narrow bar. The concentration of electric and magnetic fields reaches respectively 60 and 390 conventional units.

If there is a need to view a more complete picture of the set of resonances, you can refer to the paper \cite{sydorchuk-2017}. In this paper, the first seven resonances lowest in the frequency were studied in details including the resonant field distribution inside the unit cell of a structure  very similar to the metasurface being under our investigation here to reveal the origin of each resonance.

Thus, below we will study a nonlinear reflection by the metasurface with two different bars in the unit cell at frequencies close to the relative resonance frequency 0.6426.

\section{Nonlinear reflection regime}

\begin{figure}[]
	\centering
	\includegraphics[width=12.0cm]{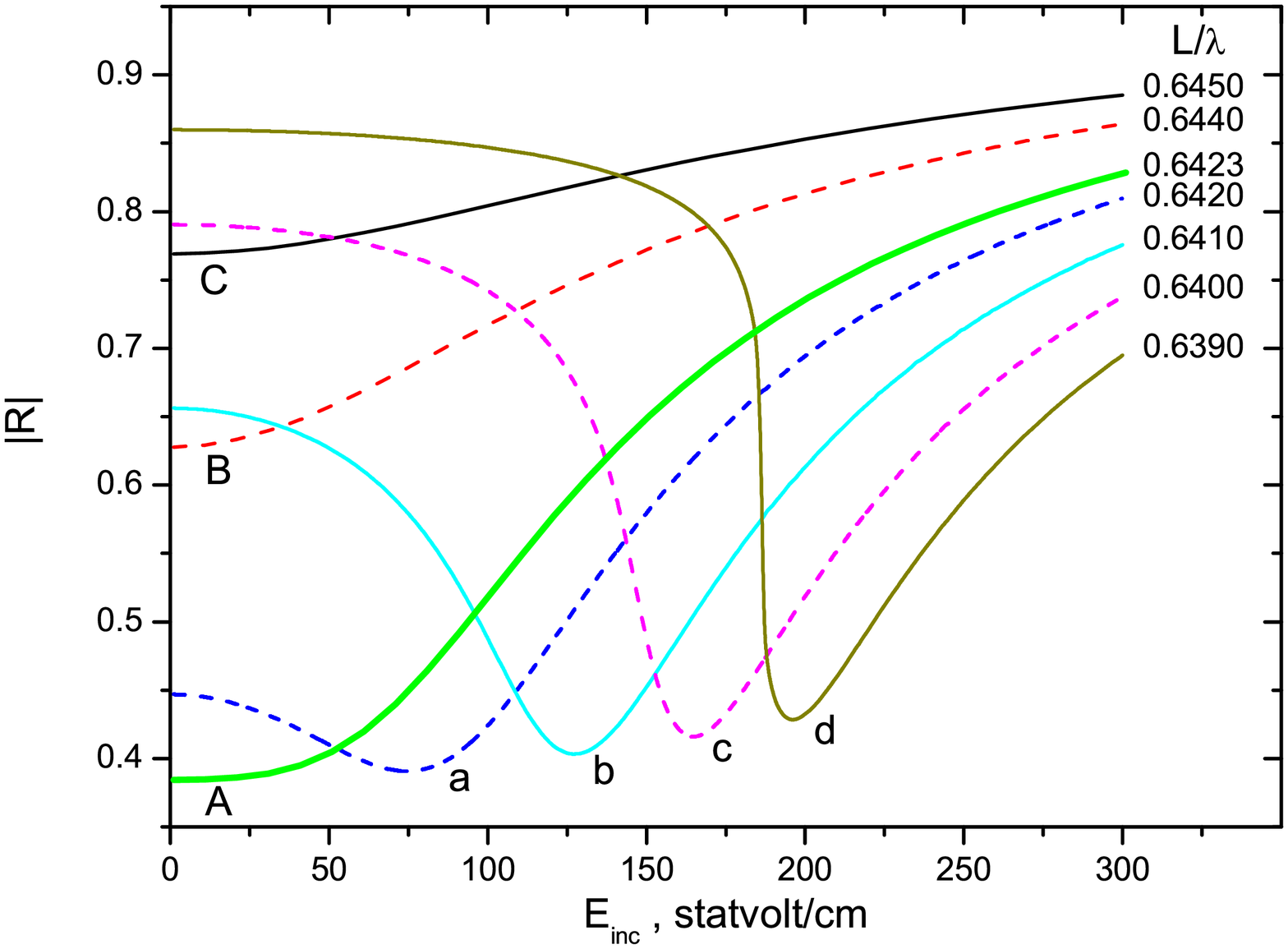}
\caption{Reflection coefficient of a normally incident plane wave at the frequencies close to the trapped mode resonance frequency as a function of the electric field strength for different relative frequency $L/\lambda$ values: A - 0.6426 (resonance frequency), B - 0.6440 and C - 0.6450 which are higher values than the resonance frequency of the linear reflection regime, and a - 0.6420, b - 0.6410, c - 0.6400, d - 0.6390 which are below the resonance frequency of reflection in the linear regime.}
\label{RfromEinc_1}
\end{figure}

\begin{figure}[]
	\centering
	\includegraphics[width=12.0cm]{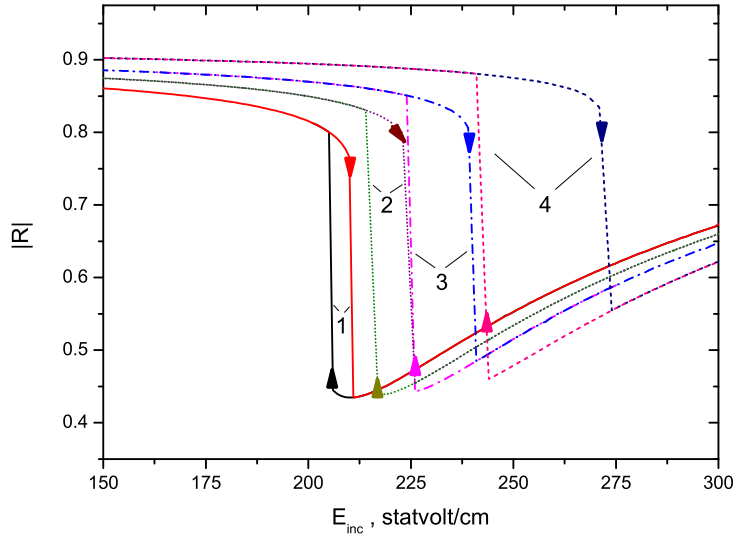}
\caption{Reflection coefficient of a normally incident plane wave at the frequencies close to the trapped mode resonance frequency as a function of the electric field strength for different relative frequency $L/\lambda$ values: 1 - 0.6385, 2 - 0.6383, 3 - 0.6380 and 4 - 0.6375. Arrows show variation of reflectivity of metasurface with increasing and decreasing of the electric field strength of the incident wave. }
\label{RfromEinc_2}
\end{figure}

In the nonlinear reflection regime the intensity of an incident electromagnetic field is large enough to produce  heating  inside the lossy resonant metasurface  that can essentially modify the permittivity of silicon bars. The effect may be interpreted as a third-order Kerr nonlinearity \cite{doi:10.1002/lpor.201700108}. As a result, the reflectivity of the structure depends on the amplitude of the incident wave and may be controllable. At a frequency close to the trapped mode resonance of the linear regime, we can observe sharp variations in dependencies of the reflection coefficient on the incident wave amplitude, Fig.~\ref{RfromEinc_1} and \ref{RfromEinc_2}.

At frequencies higher then the resonance frequency of linear reflection, increasing of the incident wave amplitude leads to destroy of the resonance and to monotonic increasing of the reflection coefficient (see Fig.~\ref{RfromEinc_1}, lines marked by letters  B and C). We observe more interesting dependence of the reflection coefficient on the electric field strength of the incident wave at frequencies below the resonance frequency of the linear regime (see Fig.~\ref{RfromEinc_1}, lines marked by letters a, b, c, and d and also Fig.~\ref{RfromEinc_2}). The lower is the frequency the more is the electric field strength of the incident wave which corresponds to the resonance.  Moreover the metasurface  reflection manifests bistability and hysteresis in dependence on the strength of the incident field.

Bistability is a manifestation of the self-action of an electromagnetic field in the nonlinear metasurface under consideration with a feedback, in which two possible stable stationary states of the field of a reflected wave correspond to a certain intensity of the incident radiation. Manifesting a hysteresis property the amplitude of the reflected wave can take one of two values. With a cyclical change in the intensity of the incident wave over a wide range, the bistable reflection metasurface operates reversibly, and the previous state of the reflection uniquely determines which of the two stable field states is realized in the reflection field.

The reason for appearance of a region of the incident radiation intensity values, for which the reflection coefficient is two-valued, is precisely the feedback in the researched nonlinear system.

\begin{figure}[!htb]
	\centering
	\includegraphics[width=12.0cm]{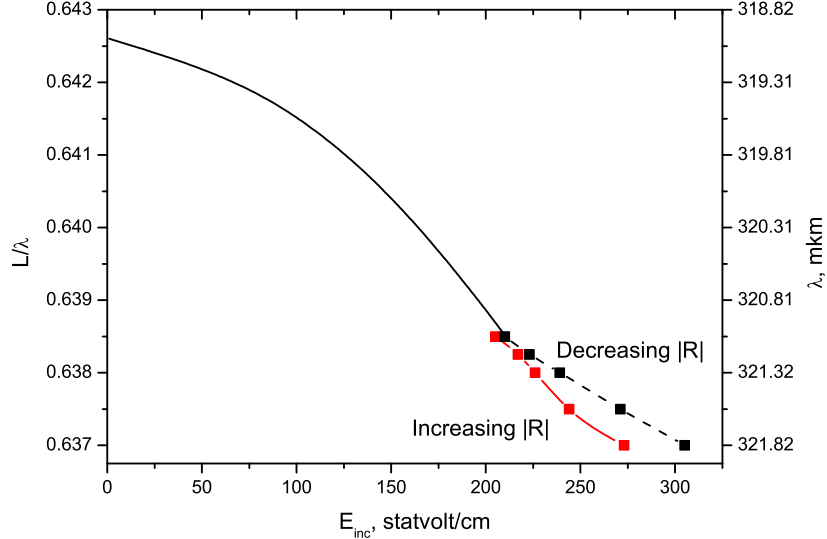}
	\caption{Dependencies of frequencies of sharp resonant increasing and decreasing of the reflection coefficient on the incident field strength.}
	\label{kappa_on_Einc}
\end{figure}

In Fig.~\ref{kappa_on_Einc} we present dependencies of frequencies of  sharp resonant increasing and decreasing of the reflection coefficient on the strength of the incident field. The hysteresis range begins from the strength  approximately equal to 200~statvolt/cm.

\begin{figure}[!htb]
	\centering
	\includegraphics[width=12.0cm]{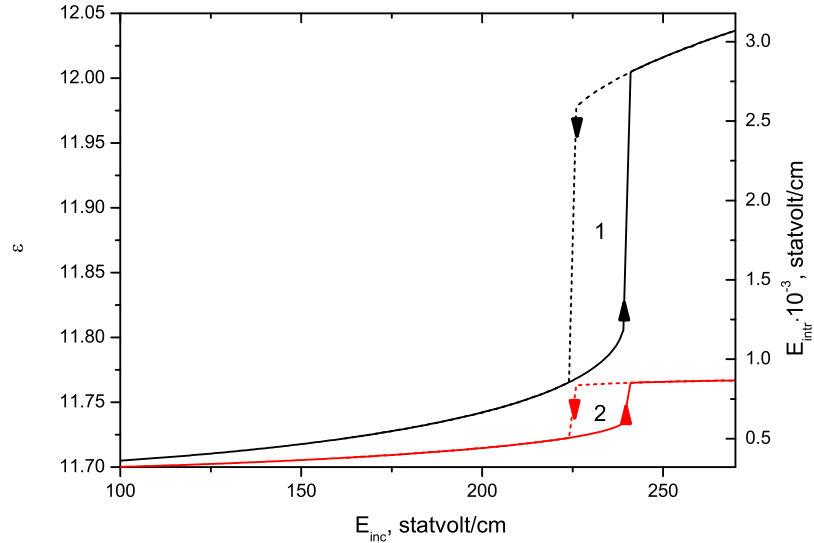}
	\caption{Variation of a permittivity and $E_{intr}$ in the middle point of each bar for the relative frequency $L/\lambda$=0.6380 with the incident field strength. Labels 1 and 2 mark the narrow and wide bar respectively.}
	\label{Eintr_eps}
\end{figure}

In the strong resonant field inside the structure, the permittivity of the semiconductor bars varies with the incident wave intensity. We have studied this variation of the permittivity in a middle point of each bar of the unit cell and presented it in Fig.~\ref{Eintr_eps}.  

\section{Conclusion}

For the first time we have studied in theory and simulated a nonlinear reflection of strong terahertz radiation from the silicon-on-metal metasurface involving  excitation of the high-Q trapped mode resonance.  Bistability and hysteresis of the response are peculiarity of the nonlinear reflection. Investigation of the internal field distribution can identify the structure areas there is a significant confinement of the electromagnetic field. In these areas, it is advisable to place active or nonlinear materials in order to obtain amplification and radiation generation. The physical effects found in the investigated structure may be useful in constructing of the metasurfaces for higher-frequency applications. The study of field distributions made it possible to classify resonances in accordance with the field structure. As a result, an opportunity appears to select the type of resonance for the sensor designing with the necessary Q-factor and the appropriate spatial distribution of the field intensity. 

\section*{Acknowledgment}

This work was supported by the National Academy of Sciences of Ukraine (Program "Basic Problems of Developing of Novel Nanomaterials and Nanotechnologies", Project 13/17-H).

\section*{References}


\bibliographystyle{unsrt}

\bibliography{NonlinearBib}

\end{document}